\pdfoutput=1
\documentclass[11pt]{article}

\usepackage{graphicx}
\usepackage{amsmath}
\usepackage{dcolumn}
\usepackage{bm}
\usepackage{slashed}
\usepackage[bookmarks,bookmarksnumbered,colorlinks=true,anchorcolor=blue,
linkcolor=blue,urlcolor=blue,citecolor=blue,breaklinks=true]{hyperref}
\usepackage[utf8]{inputenc}
\usepackage{authblk}

\usepackage{color}
\usepackage{ulem}
\newcommand{\pt}{\partial_u}

\newcommand{\be}{\begin{equation}}
\newcommand{\ee}{\end{equation}}
\newcommand{\ba}{\begin{eqnarray}}
\newcommand{\ea}{\end{eqnarray}}
\newcommand{\no}{\nonumber \\}
\newcommand{\gsim}{\mathrel{\hbox{\rlap{\lower.55ex \hbox {$\sim$}}
                   \kern-.3em \raise.4ex \hbox{$>$}}}}
\newcommand{\lsim}{\mathrel{\hbox{\rlap{\lower.55ex \hbox {$\sim$}}
                   \kern-.3em \raise.4ex \hbox{$<$}}}}

\def\roughly#1{\mathrel{\raise.3ex\hbox{$#1$\kern-.75em%
\lower1ex\hbox{$\sim$}}}}
\def\lsim{\roughly<}
\def\gsim{\roughly>}

\def\({\left(}
\def\){\right)}
\def\[{\left[}
\def\]{\right]}
\def\<{\langle}
\def\>{\rangle}
\def\pd{\partial}

\def\l{{\lambda}}

\def\d{{\delta}}

\def\o{{\omega}}

\def\e{{\epsilon}}

\def\a{{\alpha}}
\def\b{{\beta}}
\def\c{{\chi}}

\def\g{{\gamma}}
\def\h{{\eta}}

\def\n{{\nu}}
\def\r{{\rho}}

\def\t{{\tau}}

\def\ph{{\phi}}

\def\x{{\xi}}

\def\z{{\zeta}}

\def\KZ{{\text{KZ}}}

\title{\bf Kibble-Zurek Scaling in a Holographic p-wave Superconductor}

\author[2]{Yanyan Bu\thanks{yybu@hit.edu.cn}}
\author[1]{Mitsutoshi Fujita\thanks{fujita@mail.sysu.edu.cn}}
\author[1]{Shu Lin\thanks{linshu8@mail.sysu.edu.cn}}
\affil[1]{School of Physics and Astronomy, Sun Yat-Sen University, Guangzhou 519082, China}
\affil[2]{School of Physics, Harbin Institute of Technology, Harbin 150001, China}

\renewcommand\Authands{ and }

\date{\today}

\begin{document}

\maketitle

\begin{abstract}

We study the Kibble-Zurek mechanism in a 2d holographic p-wave superconductor model with a homogeneous source quench on the critical point. We derive, on general grounds, the scaling of the Kibble-Zurek time, which marks breaking-down of adiabaticity. It is expressed in terms of four critical exponents, including three static and one dynamical exponents. Via explicit calculations within a holographic model, we confirm the scaling of the Kibble-Zurek time and obtain the scaling functions in the quench process. We find the results are formally similar to a homogeneous quench in a higher dimensional holographic s-wave superconductor. The similarity is due to the special type of quench we take. We expect differences in the quench dynamics if the condition of homogeneous source and dominance of critical mode are relaxed.

\end{abstract}

\newpage

\allowdisplaybreaks

\flushbottom

\section{Introduction}

Non-equilibrium dynamics occurs ubiquitously in different physical systems. While the microscopic theories governing the dynamics can be radically different, the dynamics close enough to a critical point (second order phase transition) shows remarkable universal scaling behavior. The mechanism was first discussed in the pioneering works of Kibble and Zurek (KZ) in the context of early universe and superfluid \cite{Kibble:1976sj,Kibble:1980mv,Zurek:1985qw}. On the critical point, relaxation time of the system diverges so the system evolves non-adiabatically. It has been established by Kibble and Zurek that the system shows certain scaling behavior, which we will refer to as KZ-scaling. The KZ-scaling has been studied in diverse systems such as cold atom systems \cite{cond-mat/0610187,0807.3323,1306.4523}, heavy ion collisions \cite{1605.09341,1712.10305,1811.05081,1811.09466,1903.06075}, etc.

Experimental realization of the KZ-scaling requires tuning the system close enough to the critical point. A useful protocol is quench (thermal or quantum), in which a parameter of the system is varied in time in a designed way in order to make the system approach the critical point. Theoretical description of the critical dynamics is difficult given that a system close to the critical point is often strongly coupled and is limited to soluble models. Holography \cite{Maldacena:1997re,Aharony:1999ti} provides a useful tool for studying critical dynamics in strongly coupled regime. There have been extensive studies on the KZ-scaling of correlation function \cite{1109.3909,Basu:2012gg,1308.4061,Das:2014lda}, defect formation \cite{1406.2329,1407.1862}, entanglement entropy \cite{1702.04359,Das:2017gjy}, etc. in holographic superconductor models. However, most of previous studies focus on a s-wave superconductor, which corresponds to spontaneous breaking of a global $U(1)$ symmetry. In this paper, we consider the critical dynamics in a p-wave superconductor due to a $SU(2)$ symmetry breaking. We will find interesting similarity between the s-wave and p-wave models.

As we shall show, the p-wave superconductor model corresponds to ``model A'' of the dynamical universality class according to the classification by Hohenberg and Halperin \cite{Hohenberg:1977ym}. We then obtain on general grounds the scaling of the KZ time, the time scale for breaking-down of adiabaticity. Extending the standard Kibble-Zurek reasoning, we find the KZ time is determined by four critical exponents rather than two. This includes one dynamical and three static critical exponents. This is due to the special type of quench we use to realize the critical dynamics. We confirm the result by explicit analysis within the holographic model. We also obtain explicitly the scaling function for the condensate in the quench dynamics.

The rest of this paper will be organized as follows. In Section \ref{model_exponents} we review the basic ingredients of the holographic p-wave superconductor model, followed by the calculations of all static and dynamical critical exponents in this model.
In Section \ref{sec:KZ}, on general grounds, we express the KZ time in terms of four critical exponents, which is further confirmed by a numerical study within the holographic model. In Section \ref{KZ_bulk_eom}, we analyze the quench dynamics from the bulk equation of motion. We find a special role is played by a zero mode at exactly the critical point. The dominance of the zero mode leads to breaking-down of adiabaticity, which confirms the KZ time obtained in Section \ref{sec:KZ}. In Section \ref{KZ_condensate}, we obtain the scaling function for the condensate. The results of Sections \ref{KZ_bulk_eom} and \ref{KZ_condensate} show formal similarity with the quench study in a higher dimensional holographic s-wave superconductor. We argue that the results on the scaling of KZ time and correlation function are independent of the system's dimensionality. We conclude in Section \ref{conclusion} and discuss possible extensions of the quench dynamics considered in present work, where we do expect interesting differences with the s-wave models. Two appendices \ref{overview_exponents} and \ref{EF_KZ} present an overview of the critical exponents and derivation of the KZ-scaling in the ingoing Eddington-Finkelstein coordinates, respectively.

\section{Critical exponents of Holographic p-wave superconductor}
\label{model_exponents}

\subsection{2d holographic p-wave superconductor: overview} \label{p_model}

In recent years, an example of the p-wave superconductor called $\rm Sr_2RuO_4$ has been discovered and presents good understanding of strongly coupled electron systems~\cite{Maeno94,Maeno00}. While the Cooper pair is usually a spin singlet, the superconductor of two bound electrons with the same spin can theoretically be contemplated.  $\rm Sr_2RuO_4$ is a p-wave superconductor of a spin triplet.
In a p-wave superconductor, the spatial part of the wave-function is parity odd and the spin part turns out to be a triplet. The total wave-function describes the anti-commuting property of fermions.

Holographic models are useful for studying p-wave superconductors. In this section, we review the basic ingredients of a holographic 2d p-wave superconductor. The brane configuration such as $D3-D3'$ was used to derive the holographic 2d p-wave superconductor in \cite{Gao:2012yw,Bu:2012qr}. The number of $D3$ and $D3'$ is $N$ and 2, respectively. In the gravity dual, the probe $D3'$ branes in $AdS_5\times S^5$ realize 3d $SU(2)$ Yang-Mills theory in an $AdS_3$ black brane background. The 2d p-wave superconductor is realized with the help of the non-linear interactions of Yang-Mills theory. Since quantum fluctuations preventing the formation of the condensate are suppressed in the large $N$ limit, one can evade the Coleman-Mermin-Wagner theorem in lower dimensional theories~\cite{Anninos:2010sq}.

The back-reaction of $SU(2)$ Yang-Mills fields is important in analyzing interesting physics such as the entanglement entropy~\cite{Cai:2012nm,Arias:2012py}. However, from the viewpoint of the 10d supergravity (in which the 3d Einstein-Yang-Mills theory is supposed to be embedded), the dilaton will run when the back-reaction is taken into account. Instead, we will be limited to a toy model consisting of Einstein-Yang-Mills theory in an asymptotic $AdS_3$ \cite{Das:2017gjy}
\ba\label{IG1}
I_{G}=\dfrac{1}{2\kappa^2}\int d^3x\sqrt{-G}\Big(R+\dfrac{2}{l^2}-\dfrac{1}{g_{3}^2}\mbox{Tr}(F_{\mu\nu}F^{\mu\nu})\Big),
\ea
where $F_{\mu\nu}=\partial_{\mu}A_{\nu}-\partial_{\nu}A_{\mu}-i[A_{\mu},A_{\nu}]$  (see also higher dimensional holographic p-wave superconductor models~\cite{Gubser:2008wv,Ammon:2009xh,Cai:2015cya}). Mass dimensions of $g_{3}$ and $1/\kappa^2$ are 1.

The Einstein equation and the equation of motion (EOM) in terms of the $SU(2)$ gauge field can be derived from \eqref{IG1}:
\ba\label{EIN22}
&R_{\mu\nu}-\dfrac{1}{2}g_{\mu\nu}\Big(R+\dfrac{2}{l^2}\Big)={{\kappa}^2} T_{\mu\nu}, \nonumber \\
&\partial_{\mu}(\sqrt{-G}F^{\mu\nu})-i\sqrt{-G}[A_{\mu},F^{\mu\nu}]=0,
\ea
where $T_{\mu\nu}$ is the energy momentum tensor in the bulk. Since $\kappa$ is an overall coefficient, it decouples from the remaining parameters in the EOM \eqref{EIN22}.

First, lets consider a homogeneous background solution for the bulk theory \eqref{IG1} that depends on the radial coordinate only.
A self-consistent ansatz for the bulk metric $g_{\mu\nu}$ and gauge field $A_\mu$ is
\ba
&&ds^2=\dfrac{l^2}{z^2}\Big(-f(z)dt^2+dy^2+\dfrac{dz^2}{n(z)f(z)}\Big), \nonumber \\
&&A=\dfrac{1}{2}\Big(\sigma^1\cdot w(z) dy +\sigma^3\cdot \phi(z) dt \Big),\quad A_z^b=0,
\ea
where $\sigma^a$ ($a=1,2,3$) are Pauli matrices and $(t,y)$ are the boundary coordinates. $z$ labels the radial direction with $z=0$ the $AdS$ boundary whereas $z=z_h$ the event horizon (i.e. $f(z=z_h)=0$).
The backgrounds $f(z)$, $n(z)$, $w(z)$ and $\phi(z)$ could be derived by solving \eqref{EIN22} with regularity condition at the horizon $z=z_h$
\ba
f(z_h)=0,\quad \phi(z_h)=0, \quad n(z_h)=\text{const}, \quad w(z_h)=\text{const},
\ea
and the $AdS$ condition at the boundary $z=0$
\ba
&\phi(z) \to  -\rho \log z+\mu, \quad w(z)\to  -J \log z+w_v, \quad {\rm as}~ z\to 0,
\ea
where $\mu$ and $\rho$ correspond to the chemical potential and charge density, respectively.

With a vanishing source $J=0$, there are two solutions for the bulk theory \cite{Gao:2012yw,Bu:2012qr,Das:2017gjy}: a charged $AdS_3$ black brane without a hair and a hairy $AdS_3$ black brane. The former corresponds to the normal phase ($w_v=0$) of the p-wave superconductor and is thermodynamically favorable in the high temperature regime. When the temperature becomes lower than a critical value, the hairy $AdS_3$ black brane is more stable and corresponds to the superconductor phase ($w_v\neq 0$). As the order parameter of the superconducting phase transition, $w_v$ is encoded in the near-boundary behavior of bulk gauge field $w(z)$. Note the hairy solution with a vector hair corresponds to a solution signifying a spontaneous symmetry breaking with broken parity. Within such a holographic system, the gap formation, AC conductivity and zero modes were considered in the condensed phase in the probe limit \cite{Gao:2012yw,Bu:2012qr}. 
In \cite{Das:2017gjy}, the entanglement entropy of this holographic model was shown to bear a non-monotonic behavior depending on subsystem size. The non-monotonic behavior is also expected in other models of superconductivity (e.g. 2d s-wave superconductor) and superfluidity. This is caused by a competition between the formation of the condensate and the effect of charged density.



In the normal phase, the component $w(z)=0$ which reduces the bulk theory \eqref{IG1} into the Einstein-Maxwell theory in asymptotic $AdS_3$.
%
%
Thus, the charged $AdS_3$ black brane solution without a hair is obtained as~\cite{Banados:1992wn,Cadoni:2009bn}\footnote{One can use the rescaling of $g_3$, because the EOM depends only on $g_3l$.}
\ba\label{btz310}
ds^2_{BTZ}=\dfrac{l^2}{z^2}\left(-f(z)dt^2+dy^2+\dfrac{dz^2}{f(z)}\right), \quad A_t^3=\phi = \rho \log \left(\dfrac{z}{z_h}\right),
\ea
where $f(z)=1-(z/z_h)^2+(\rho z)^2/\tilde{g}_{3}^2\log (z/z_h)$.
The Hawking temperature is
\ba
T_H=\dfrac{1}{4\pi}\left(\dfrac{2}{z_h}-\dfrac{\rho^2 z_h}{\tilde{g}_3^2}\right).
\ea

In the normal phase, the free energy has been analytically obtained in~\cite{Jensen:2010em,Fujita:2018xkl}. After adding the Gibbons-Hawking term and counter terms to the action \eqref{IG1}, one derives a finite action. The free energy in the normal phase is
\ba
F/V_1=\dfrac{1}{2\kappa_3^2}\left(-\dfrac{1}{2z_h^2}+\dfrac{q^2 (1+\log (z_h/l))}{2\tilde{g}_{3}^2}\right),
\ea
where we have introduced dimensionless parameters $\kappa_3^2=\kappa^2/l$, $\tilde{g}_3=g_3 l$, and $z_h=1/(\pi T_H+\sqrt{\frac{q^2}{2\tilde{g}_{3}^2}+\pi^2 T_H^2})$.
The specific heat is computed as
\ba\label{SPE9}
 C_H=-T_H\dfrac{\partial^2 F}{\partial T_H^2}=T_H\dfrac{\partial S}{\partial T_H} =\dfrac{{\pi^2} T_HV_1} {\kappa_3^2} \left(1+ \dfrac{\sqrt{2}\pi} {\sqrt{2\pi^2+ (q/(\tilde{g}_{3}T_H))^2}}\right)>0,
\ea
which is always positive except for the case of zero temperature.


\subsection{Determination of static critical exponents}
In this section, we obtain the static critical exponents of the holographic p-wave superconductor reviewed in section \ref{p_model}.
In appendix \ref{overview_exponents}, we present a review of the critical exponents, which include six static and one dynamical exponents.

The exponent $\alpha$ reveals the power behavior of the specific heat: $C_H\sim |\epsilon_T|^{-\alpha}$, where $\epsilon_T=T/T_c-1$ and $T_c$ is critical temperature. Focusing on the normal phase, we find from \eqref{SPE9} that the specific heat converges to a constant value and gives the exponent $\alpha =0$. The specific heat in the condensed phase can be computed numerically \cite{Das:2017gjy}. It presents the same value for the exponent $\alpha$.

The exponent $\beta$ denotes the power of the order parameter in the condensed phase: $w_v\sim |\epsilon_T|^{\beta}$. Its value has been derived in \cite{Gao:2012yw,Bu:2012qr} as $\beta =1/2$.

For the rest four static exponents, one can obtain them through direct computations using the definitions summarised in the appendix \ref{overview_exponents}. However, they obey hyperscaling relations \eqref{RUS62}-\eqref{JOS65}, which have also been confirmed in \cite{Maeda:2009wv} in s-wave holographic superconductor models. We assume the hyperscaling relations also hold in our background. Consequently, we obtain all the six static exponents:
%
%
\ba\label{ce_static}
(\alpha,\ \beta,\ \gamma, \ \delta, \ \eta, \ \nu)=\left(0,\ \frac{1}{2}, \ 1,\ 3,\ 1,\ 1\right).
\ea
Obviously, the static exponents of the holographic p-wave superconductor are of the mean field type \eqref{ce_mf}.

\subsection{Dynamical critical exponent from perturbation of condensate}

For the dynamical exponent, we follow~\cite{Maeda:2009wv} (see also \cite{1703.00933,1807.11881}) and analyze correlation functions of the vector condensate. Our approach is based on the normal phase at high temperature because one can use the analytical charged black hole solution \eqref{btz310}, which does simplify the algebras a lot.
For later convenience, we perform the rescaling $(t,\ y,\ z) \to (z_h t,\ z_h y, \ z_h z)$ and $\rho \to \rho/z_h$ in \eqref{btz310}.

We consider the perturbation around the charged black hole solution \eqref{btz310} $A_{\mu}=A_\mu+\delta a_\mu$.
The perturbation has the Fourier exponent $\delta a_{\mu}=\sum_a \delta a^{a}_{\mu}(z)\sigma^a e^{-i\omega t+i q y}/2$, where $\delta a^1_{\mu}=\delta w_{\mu}(z)$, $\delta a^2_{\mu}=\delta g_{\mu}(z)$, and $\delta a^3_{\mu}=\delta \phi_{\mu}(z)$. We use the radial gauge $\delta a_z=0$. When $q=0$, we have the following two decoupled sectors: $\{\delta a_t^1,\ \delta a_t^2,\ \delta a_y^3\}\equiv\{ \delta w_t,\   \delta g_t,\ \delta \phi_y \}$ and $\{a_y^1,\ a_y^2,\ a_t^3\}\equiv\{ \delta w_y,\   \delta g_y,\ \delta \phi_t \}$, satisfying the following EOMs:
\ba
\dfrac{\phi (z) w(z) \delta \phi_y}{f}+\dfrac{\delta w_t'}{z}+\delta w_t''&=&0, \nonumber \\
-\dfrac{i\omega w(z)\delta \phi_y}{f(z)}-\dfrac{w(z)^2\delta g_t }{f}+\dfrac{\delta g_t'}{z}+\delta g_t''&=&0, \nonumber \\
\dfrac{-i\omega w(z)\delta g_t-\phi(z) w(z)\delta w_t}{f^2}+\dfrac{\omega^2\delta \phi_y}{f^2}+\left(\dfrac{f'}{f}+\dfrac{1}{z}\right)\delta \phi_y'+\delta\phi_y''&=&0, \nonumber \\
\phi'(z) \delta g_t-\phi (z)\delta g_t'-i\omega\delta w_t'&=&0, \nonumber \\
f \left(w(z) \delta \phi_y'-w'(z) \delta \phi_y \right)+\partial_t \delta g_t'-\phi'(z) \delta w_t+\phi(z) \delta w_t'&=&0,
\ea
and
\ba 
\dfrac{i\omega w(z) \delta g_y-w(z)^2 \delta \phi_t-2 \phi (z) w(z) \delta w_y}{f}+\dfrac{\delta\phi_t'}{z}+\delta \phi_t''&=&0, \label{del_phit} \\
\dfrac{2 \phi(z) \left(-i\omega \delta g_y+w(z)\delta \phi_t \right)+\phi(z)^2 \delta w_y+\omega^2 \delta w_y}{f^2} +\dfrac{\delta w_y'}{z}+\dfrac{f'(z) \delta w_y'}{f}+\delta w_y''&=&0, \label{del_wy} \\
\dfrac{\phi (z)^2 \delta g_y+\omega^2 \delta g_y+i\omega w(z) \delta \phi_t+2i\omega \phi (z)\delta w_y}{f^2} +\dfrac{f'(z) \delta g_y'}{f}+\dfrac{\delta g_y'}{z}+\delta g_y''&=&0, \label{del_gy} \\
f \left(w'(z) \delta g_y-w(z) \delta g_y'\right)-i\omega \delta \phi_t'&=&0. \label{del_gy_phit}
\ea

We are interested in the fluctuations of the vector condensate, which correspond to the sector $\{ \delta w_y,\   \delta g_y,\ \delta \phi_t \}$.
%
%
%
Moreover, in the normal phase, $w(z)$ vanishes so that $\delta \phi_t$ turns out to be trivial and could be set to zero (from \eqref{del_phit}). Then, the constraint equation \eqref{del_gy_phit} also becomes trivial. As a result, in terms of the combinations $\Phi_{\pm} \equiv \delta w_y\pm i\delta g_y$, the rest equations \eqref{del_wy} and \eqref{del_gy} get into decoupled forms:
\ba
&\Phi_+''(z)+\dfrac{\left(z f(z) f'(z)+f(z)^2\right) \Phi_+'(z)}{z f(z)^2}+\dfrac{\Phi_+(z) (\omega-\rho  \log (z))^2}{f(z)^2}=0, \label{Phi_+}\\
&\Phi_-''(z)+\dfrac{ \left(z f(z) f'(z)+f(z)^2\right)\Phi_-'(z)}{z f(z)^2}+\dfrac{\Phi_-(z) (\omega+\rho  \log (z))^2}{f(z)^2}=0. \label{Phi_-}
\ea
Near the $AdS$ boundary $z=0$, the solutions for $\Phi_{\pm}$ behave as
\ba
&\Phi_{\pm}=\Phi_{\pm}^{(0)}+\Phi_{\pm}^{(1)}\log z+\cdots.  \nonumber
\ea
Near the horizon $z=z_h$, we impose the ingoing-wave boundary condition so that $\Phi_{\pm}\sim (1-z^2)^{-i\omega /2}$, which denotes that waves entering the black hole horizon can not escape from the interior.

When $\rho=q=0$, there exist analytic solutions~\cite{Ren:2010ha,Zeng:2012xy}
\ba \label{analytic_Phi}
\Phi_{\pm}=(1-z^2)^{-\frac{i\omega}{2}}{}_2F_1 \Big(1-\dfrac{i\omega}{2},-\dfrac{i\omega}{2},1-i\omega, 1-z^2\Big).
\ea
The two point correlation function is given by $G(\omega, q)=-\Phi_{\pm}^{(0)}/\Phi_{\pm}^{(1)}$. With the analytic solutions \eqref{analytic_Phi}, the correlation function is
\ba
G(\omega, 0)=-\Big(\gamma + \dfrac{i}{\omega}+\psi \Big(-\dfrac{i\omega}{2}\Big)\Big).
\ea
In the hydrodynamic limit, $G(\omega, 0)$ is expandable
\ba
G(\omega, 0)=\dfrac{i}{\omega}+\dfrac{i}{12}\pi^2\omega+\dfrac{\psi (2,1)}{8}\omega^2+\cdots .
\ea
The above correlation function is divergent like $1/\omega$ at $\omega =0$.

When $\rho\neq 0$, we consider the probe limit $g_3\gg 1$. Then, $T_{\mu\nu}$ could be ignored on the right-hand side of \eqref{EIN22} and the charged $AdS_3$ metric in \eqref{btz310} could be replaced by an $AdS_3$ black hole. The probe limit makes the analysis simpler and is also well-defined in string theory~\cite{Gao:2012yw}.  The critical point is known as the point where $\Phi_{\pm}^{(1)}$ becomes zero when $\omega=0$, which, in terms of $\rho$, is $\rho_c=21.7T_H$. The deviation from the critical point could be illustrated by $\epsilon_\rho =(\rho-\rho_c)/\rho_c$, which is equivalent to $\e_T=(T-T_c)/T_c$.

\begin{figure}[htbp]
     \begin{center}
          \includegraphics[height=5cm,clip]{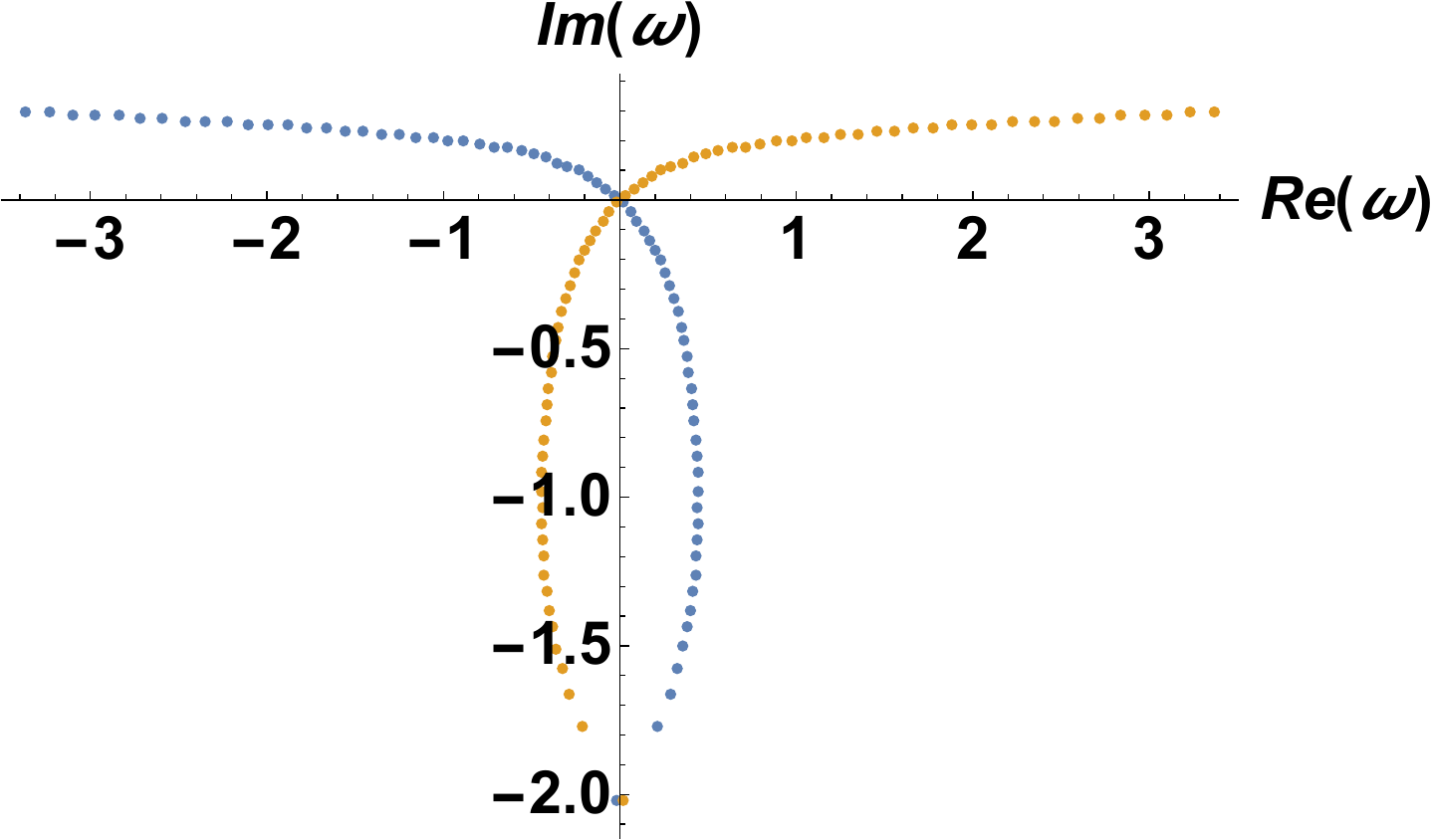}
          \hspace{1.6cm}
   \includegraphics[height=5cm,clip]{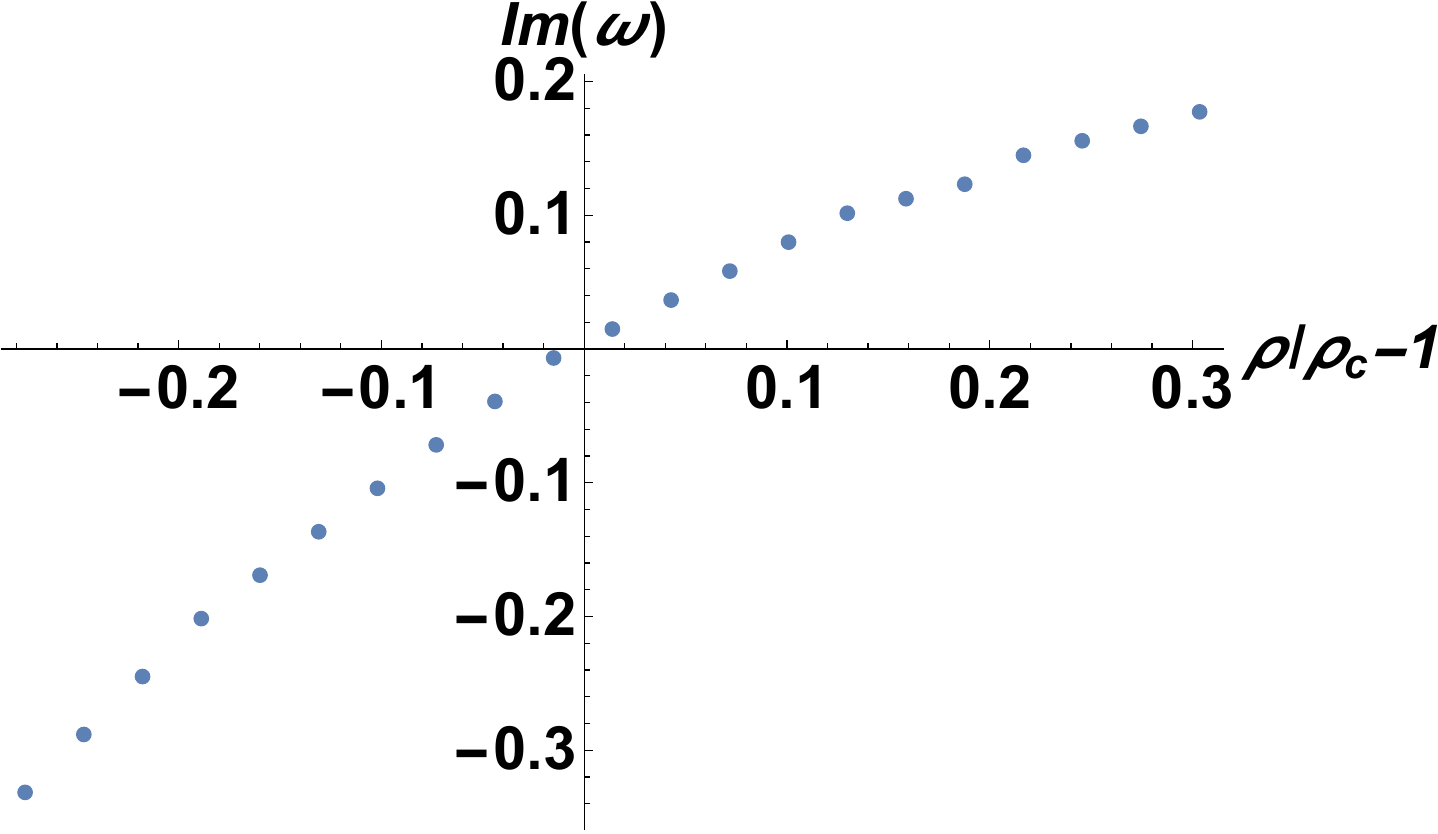}
           \caption{Top: The lowest QNM of $\Phi_+$ (blue line) and $\Phi_-$ (orange line) as $\rho$ is varied. The QNM is located at $-2i$ when $\rho =0$. It goes across the origin at $\rho=\rho_c$.
Bottom: The imaginary part of the lowest QNM as a function of $\epsilon_{\rho}$. It goes to the origin on the critical point.   }
    \label{fig:QN}
    \end{center}
\end{figure}

Solving the fluctuation equations \eqref{Phi_+},\eqref{Phi_-} with ingoing wave condition at the horizon and Dirichlet boundary condition $\Phi_\pm^{(1)}=0$ at the boundary, we obtain the spectrum of the quasi-normal modes (QNMs).
The behavior of the lowest QNM is plotted in Figure. \ref{fig:QN}, which explicitly demonstrates that the lowest QNM goes to the origin at exactly the critical point $\rho =\rho_c$ as a function of $\epsilon_{\rho}$. As a function of $\epsilon_{\rho}$, the imaginary part of the lowest QNM behaves as
\ba\label{QNM323}
\mbox{Im}(\omega_{QNM})=0.0009175+0.89666 \epsilon_{\rho}-1.08382 \epsilon_{\rho}^2\propto \epsilon_{\rho}.
\ea

The imaginary part of the lowest QNM is interpreted as the inverse relaxation time due to  $\exp (-i \omega t)\sim \exp (\mbox{Im}\omega t)\sim \exp (- t/\tau)$.
The linear fit of $\mbox{Im}(\o_{QNM})$ near the critical point indicates that $\t\sim\e_\r$. On the one hand, the critical scaling of relaxation time is given by $\t\sim \x^{z}$ with $z$ the dynamical critical exponent\footnote{This $z$ for the dynamical critical exponent should not be confused with the radial coordinate $z$ in the holographic model.}; on the other hand, the critical scaling of the correlation length $\x\sim \e_\r^\n$ with $\n=1$ from \eqref{ce_static}. The dynamical critical exponent is then given as $z=1$. It is consistent with the prediction from Hohenberg and Halperin \cite{{Hohenberg:1977ym}}: as the order parameter is non-conserved and does not couple to the stress tensor, the holographic system corresponds to model A with $z=2-\h$.

\section{KZ time from adiabaticity break-down}\label{sec:KZ}

\subsection{KZ time in terms of critical exponents}

In this section, on general grounds, we extend the original KZ reasoning to the case of a source quench.
Near the critical point (second order phase transition), both the correlation length and relaxation time diverge:
\begin{align}\label{scaling}
&\x\propto |\e_T|^{-\n},
&\t\propto \x^{z}\propto |\e_T|^{-z\n}.
\end{align}
We are interested in a black hole background, for which the relaxation is well defined and
corresponds to dissipation in the black hole background\footnote{In the case of soliton background, the counterpart is the change rate of energy spectrum: $E\frac{dE}{dt}$ where $E$ is the ground state energy.}.
Consider a homogeneous source quench with the time dependence $J\sim vt$ applied to the system on precisely the critical point, where the relaxation time is infinite. Following the KZ reasoning, adiabaticity is lost when the time to critical point is comparable to the corresponding relaxation time:
\begin{align}\label{KZ_arg}
t\sim\t \sim|\e_T|^{-z\n}.
\end{align}
We still need to express the deviation from the critical point $\e_T$ by the source $J$. This is where the other two critical exponents enter. Note that the source $J$ and the corresponding vector condensate $w_v$ are mapped to external magnetic field $h$ and the magnetization $m$ in a ferromagnetic phase transition. We can obtain the dependence of $\e_T$ on $J$ by the following scaling relations from \eqref{delta_eta} and \eqref{beta}:
\begin{align}
w_v\sim|\e_T|^\b,\quad w_v\sim J^{\frac{1}{\d}}.
\end{align}
Identifying the states with the same vacuum expectation value (VEV), we obtain $|\e_T|\sim J^{\frac{1}{\d\b}}$, and consequently
\begin{align}\label{tauJ_sacling}
\t\sim J^{-\frac{z\n}{\d\b}}\equiv J^{-\z}.
\end{align}
Our explicit results on critical exponents in the holographic p-wave model gives $\z=2/3$. 
The KZ time is then obtained from \eqref{KZ_arg} as
\begin{align}\label{KZ_time}
t_{\KZ}\sim\t\sim(vt_{\KZ})^{-\z},\quad \Rightarrow t_{\text{KZ}}\sim v^{-\frac{\z}{\z+1}}\sim v^{-2/5}.
\end{align}

\subsection{A critical exponent with the source $J$}
As an independent check, we also verify \eqref{tauJ_sacling} by a numerical study of the QNM with a {\it staic} source $J$ on the critical point. We derive a critical exponent as a function of the source $J$ like \eqref{tauJ_sacling}. 
The {\it static} source $J$ perturbs the system away from the critical point. It induces static response of charge density and condensate. In our model, this corresponds to static profile of $A_y^1$ and $A_t^3$. We then consider fluctuation of $A_y^2$ in this background and look for its lowest QNM, which gives the relaxation time of the system away from critical point. With this mind, we assume the ansatz for fields $A_y^1=w(z)$, $A_t^3=\phi(z)$, and $A_y^2=g(z)\exp(-i\omega t)$.
Dropping the backreaction of $g$ to $w$ and $\ph$, we obtain the following EOM.
\begin{align}\label{EOM22}
&z \phi''+\phi'-\dfrac{z \phi w^2}{f}=0, \nonumber  \\
& z f w''+\left(1-3 z^2\right) w'+ \dfrac{z \phi^2 w}{f}=0, \nonumber  \\
& z f g''+\left(1-3 z^2\right) g' +\dfrac{z \left(\omega^2+\phi^2\right)g}{f}=0.
\end{align}
Note that \eqref{EOM22} is to all order in $w$ and $\ph$, but only linear in $g$, whose QNM we now solve for numerically.
We require the regularity boundary condition for $\phi$ and $w$ at the black hole horizon as follows:
\ba\label{HOR23}
&\phi\sim a (z-1)+O((z-1)^2),\quad w\sim b+O(z-1).
\ea
In addition, we impose the ingoing-wave boundary condition for $g$ at the black hole horizon $g\sim (1-z)^{-\frac{i\omega}{2}}$ with an overall coefficient one.
Near the $AdS$ boundary, the fields are expanded as
\ba
w\sim -J\log z +w_v,\quad \phi\sim -\rho \log z+\mu, \quad g\sim -g_1\log z+g_2.
\ea
Solving the EOM \eqref{EOM22} and using parameters of the horizon expansion \eqref{HOR23}, we derive $J$ and $\rho$ on the boundary. We need to tune the horizon parameters such that the system remains on the critical point $\r=\r_c=3.45$ for varying $J$. 

\begin{figure}[htbp]
     \begin{center}
   \includegraphics[height=5cm,clip]{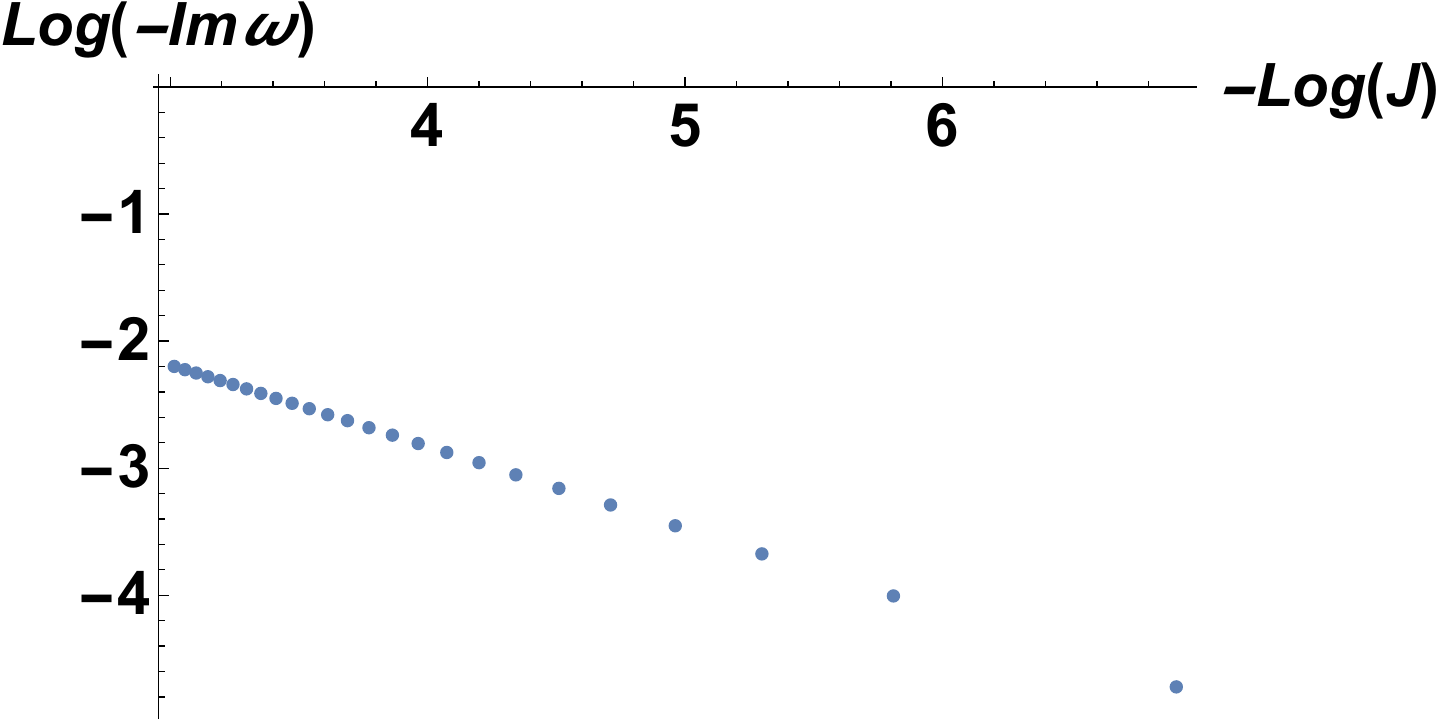}
           \caption{$\log (-\mbox{Im}\omega)$ as a function of $-\log J$. The minus sign demonstrates that $\mbox{Im}\omega$ is negative in the high temperature phase. The value of the slope is $-0.65$, which is consistent with $2/3$ from our derivation on general grounds. \eqref{tauJ_sacling}}
    \label{fig:ImJ}
    \end{center}
\end{figure}

The QNM frequencies are numerically determined from a zero of $|g_1/g_2|$. We obtain a QNM frequency near the zero of the complex plane and the critical point $\rho=\rho_c$ (in the normal phase). The real part of this QNM frequency is almost zero with changing of $J$. In Figure. \ref{fig:ImJ}, the imaginary part is plotted as a function of $-\log J$. The value of the slope is -0.65, which is consistent with $2/3$ derived on general grounds \eqref{tauJ_sacling}. This implies that the relaxation time has the scaling in terms of the source $J$ as follows:
\ba
-\mbox{Im}\omega \sim \dfrac{1}{\tau} \sim J^{\frac{2}{3}},
\ea
where the minus sign is due to the negative value of the QNM frequencies $\omega$. The negative QNM frequencies denote the stability of the system in the normal phase. 

\section{Kibble-Zurek time from bulk EOM} \label{KZ_bulk_eom}

We wish to confirm the breaking down of adiabaticity from analysis of bulk EOM in the p-wave background. In the probe limit, a consistent ansatz for the bulk gauge fields is 
\begin{align}
A_t^3=\phi(t,z),\quad A_y^1=w(t,z),\quad A_y^2= g(t,z),\quad {\rm others}=0
\end{align}
The EOMs are given by
\begin{align}\label{EOM11}
&-(\ph'+z\ph'')+\frac{z}{f}\(g^2\ph+w(w\ph+\dot{g})-g\dot{w}\)=0, \no
&zw\ph^2+zff'w'+f^2\(w'+zw''\)+2z\ph\dot{g}+z\(g\dot{\ph}-\ddot{w}\)=0, \no
&zg\ph^2+zff'g'+f^2\(g'+zg''\)-2z\ph\dot{w}-z\(w\dot{\ph}+\ddot{g}\)=0,
\end{align}
%
%
where the dot and prime represent derivative with respect to $t$ and $z$, respectively. In \eqref{EOM11}, we do not keep the constraint equation, which, once satisfied by the initial condition, holds automatically. It is interesting to note that the structure of \eqref{EOM11} is formally similar to that of the holographic s-wave model in \cite{Das:2014lda} provided that we identify $\ph$ with the Maxwell field, $w$ and $g$ with the real and imaginary parts of the charged complex scalar.
We analyze the evolution of the fields subject to the external source $J$ on precisely the critical point, when the system is just about to condense. The background (initial configuration) is given by
\begin{align} \label{initial}
\ph=\ph_0(z)=-\r_c\ln z,\;\; w=0,\;\; g=0.
\end{align}
We activate a small source $J=J_0\tanh(vt)$ with $J_0\ll 1$. At $|t|\gg 1/v$, the source can be considered as adiabatic. We can do an adiabatic expansion of the fields
\begin{align}\label{adiabatic}
&\ph=\ph_0(t,z)+\e \ph_1(t,z)+\cdots, \no
&w=w_0(t,z)+\e w_1(t,z)+\cdots, \no
&g=\e g_1(t,z)+\cdots.
\end{align}
$\e$ is a book-keeping parameter counting number of time derivatives in the source $J$ entering the fields. The leading order results contain no time derivative:
\begin{align}\label{LO}
\ph_0=-\r_c\ln z+J(t)^{2/3}\a(z),\quad w_0=-J(t)F(z)+J(t)^{1/3}\c(z),\quad g_0=0,
\end{align}
with $F\sim\ln z$, $\c\sim1$ near the boundary corresponding to source and VEV respectively.
Note that the fields are perturbed by the source adiabatically. The appearance of the fractional powers is closely tied to the existence of a zero mode $\c(z)$ on the critical point
\begin{align} \label{D_operator}
D\c(z)\equiv \frac{z\r_c^2\ln^2z}{f}\c(z)+\(\pd_z \(zf\pd_z\)\)\c(z)=0.
\end{align}
The next to leading order corrections to the fields come from time derivative of $J$, or equivalently time derivatives of $\ph_0$ and $w_0$. They satisfy the following EOMs
\begin{align}\label{EOM_NL}
&-\ph_1'-z\ph_1''+\frac{2zw_0w_1\ph_0}{f}+\frac{zw_0^2\ph_1}{f}=0, \no
&zw_1\ph_0^2+2zw_0\ph_0\ph_1+f^2w_1'+zff'w_1'+zf^2w_1''=0, \no
&zg_1\ph_0^2+f^2g_1'+zff'g_1'+zf^2g_1''-2z\ph_0\dot{w_0}=0.
\end{align}
It is easy to see $\ph_1=w_1=0$ is a consistent solution to \eqref{EOM_NL}. $g_1$ needs to be solved from the last equation of \eqref{EOM_NL}. To solve for $g_1$, we note that it satisfies an inhomogeneous equation and the boundary conditions $g_1\sim 1$ near the boundary and is ingoing near the horizon. We can decompose it with eigenfunctions of the operator $D$: $g_1(t,z)=\sum_na_n(t)\varphi_n(z)$ with $D\varphi_n(z)=\l_n\varphi_n(z)$. Note that we have chosen $D$ to be self-adjoint. The orthogonality condition for its eigenfunctions readily follows
\begin{align}\label{ortho}
\int dz \varphi_n(z)\varphi_m(z)=\d_{nm}.
\end{align}
We have assumed a discrete eigenvalue spectrum.
Plugging the decomposition into \eqref{EOM_NL} and keeping to the leading order in $J$, we obtain 
\begin{align}
\sum_n\[D a_n(t)\varphi_n(z)+\(-2J^{2/3}\r_c\ln z\a\)a_n(t)\varphi_n(z)\]=\r_c\ln z\dot{J}J^{-2/3}\c(z).
\end{align}
Applying $\int dz\varphi_m(z)$ on both sides and using orthogonality condition, we have
\begin{align}
\sum_n\[\l_n\d_{nm}a_n(t)+J^{-2/3}A_{nm}a_n(t)\]=B_m\dot{J}J^{-2/3},
\end{align}
with
\begin{align}
&A_{nm}=\int dz \varphi_n(z)\varphi_m(z)(-2\r_c\ln z\a), \no
&B_m=\int dz \varphi_m(z)\r_c\ln z\c(z).
\end{align}
Due to the presence of zero mode $\l_0=0$, we easily obtain the leading order solution in $J$
\begin{align}
a_0(t)=\dot{J}J^{-4/3}B_0/A_{00},\;\; a_n(t)=J^{-2/3}\dot{J}B_n/\l_n,\;\;(n\ne0).
\end{align}
This shows the solution is dominated by zero mode, from which we have $g_1\sim\dot{J}J^{-4/3}$. Adiabaticity breaks down when $g_1\sim w_0$, which leads to the condition $\dot{J}\sim J^{5/3}$ and consequently $t_{\KZ}\sim v^{-2/5}$. This agrees perfectly with the expectation \eqref{KZ_time} on general grounds.
In appendix \ref{EF_KZ}, we extend the analysis to second order using the ingoing Eddington-Finkelstein coordinates, which also confirms the KZ time.

\section{Adiabaticity Breaking-down and KZ Scaling Function} \label{KZ_condensate}
As seen in section \ref{KZ_bulk_eom}, the adiabaticity breaks down in the regime $v t\ll 1$ and then $\tanh (vt)\to vt$, where the source and the VEV scale as
\ba
J(t_{KZ})\sim J_0v^{\frac{3}{5}},\quad w_v(t_\KZ ) \sim J(t_\KZ)^{\frac{1}{3}} \sim  v^\frac{1}{5},\quad \mu(t_{\KZ})  \sim J(t_\KZ)^{\frac{2}{3}} \sim  v^\frac{2}{5}.
\ea
These scaling behaviors present an insight of the scaling behavior in the critical region.
Alternatively, we consider an expansion in terms of fractional powers of $v$ near the critical point.
Scaling relations suggest rescaling the time and fields as follows:
\ba
t\to \eta =v^{\frac{2}{5}} t,
\ea
and
\ba
&w=-J(\eta v^{-\frac{2}{5}}) u_f(z)+v^{\frac{1}{5}}u_\chi (\eta, z),\quad g=v^{\frac{1}{5}}u_g(\eta, z), \nonumber \\
&\phi= -\rho \log z+v^{\frac{2}{5}}u_\alpha (\eta, z),
\ea
where the source term $J(t)$ is separated from the remaining term. The dependence on the new time $\eta$ is included in $u_\chi$, $u_g$, and $u_\alpha$, respectively. The $AdS$ boundary behaviors are $u_f\sim \log z$, $u_\chi\sim 1$, and $u_g \sim 1$.
Note that
\ba
J(\eta v^{-\frac{2}{5}})\sim J_0v^{\frac{3}{5}}\eta.
\ea

The EOM \eqref{EOM11} can be expanded in powers of $v^{\frac{2}{5}}$ as
\ba\label{EOM67}
&v^{\frac{1}{5}} f(z)D u_\chi+v^{\frac{3}{5}} \Big(-2 \rho  z\log z \partial_\eta u_g \nonumber \\
&-2 \rho  z\log z u_{\alpha } u_{\chi }-J_0\eta f(z)D u_f\Big)+O(v)=0, \nonumber \\
&v^{\frac{1}{5}} f(z) D u_g+v^{\frac{3}{5}} \Big(-2 \rho  z\log z u_g u_{\alpha }+2 \rho  z \log z \partial_\eta u_{\chi } \Big)+O(v)=0, \nonumber \\
&v^{\frac{2}{5}} \Big(-f(z)D_\alpha u_{\alpha } -z\rho \log z (u_g^2 +u_{\chi }^2\Big)\Big)+O(v^{\frac{4}{5}})=0, \nonumber \\
&v^{\frac{2}{5}} \Big(-f(z) u_g' u_{\chi }+f(z) u_g u_{\chi }'\Big) \nonumber \\
&+v^{\frac{4}{5}} \Big(f(z)(- u_f' u_g+ u_f u_g')+\partial_{\eta } u_{\alpha }'\Big)+O(v^{\frac{6}{5}})=0,
\ea
where $D_\alpha u_\alpha =(zu_\alpha')'$ and $D$ is  defined in \eqref{D_operator}. 

The fourth equation of \eqref{EOM67} is the constraint. When it is satisfied at a constant $z$, it is also satisfied at all $z$. We require the constraint at small $z$ near the $AdS$ boundary and use $AdS$ boundary behaviors of $u_g,\ u_\chi,$ and $u_\alpha$. In the fourth equation of \eqref{EOM67}, the subleading term gives the additional equation $u_f'u_g=\partial_{{\eta}}u_\alpha'$ in the leading order of the small $z$ expansion.

First, we consider the third equation of \eqref{EOM67}.
Since $D_\alpha u_\alpha$ does not have a zero mode, $u_\alpha$ is given by
\ba\label{EOM68}
u_\alpha =-D^{-1}_\alpha \Big(\dfrac{z\rho \log z (u_g^2+u_\chi^2)}{f(z)}\Big).
\ea

We turn to the first and second equations of \eqref{EOM67}.  Since we consider the critical point $\rho=\rho_c$, $D$ has a zero mode. As done in section \ref{KZ_bulk_eom}, we decompose the fields in terms of the eigenfunctions of $D$ as follows:
\ba
D\varphi_n (z)=\lambda_n \varphi_n (z), (\lambda_0=0,\ \lambda_n>0 \ (n\ge1))
\ea
and
\ba\label{DEC43}
&u_g=\sum_n u_{g,n}(\eta )\varphi_n (z), \nonumber \\
&u_\chi=\sum_n u_{\chi ,n}(\eta )\varphi_n (z), \nonumber \\
&u_\alpha =\sum_nu_{\alpha, n}(\eta )\varphi_n (z).
\ea
Eigenfunctions satisfy the orthogonality condition.

By substituting \eqref{DEC43} into the EOM \eqref{EOM67} and defining
\ba
&\mathcal{A}_{kn}=\int \dfrac{2\rho z\log z \varphi_k^*\varphi_n}{f(z)} dz, \nonumber \\
&\mathcal{B}_{kmn}=\int \dfrac{2 \rho z\log z \varphi_k^*\varphi_m\varphi_n}{f(z)} dz, \nonumber \\
&\mathcal{J}_k=\int dz \varphi_k^*D u_f,
\ea
the first and second equations of \eqref{EOM67} are rewritten as the following infinite set of ODEs:
\ba
&\lambda_k u_{\chi, k}-v^{\frac{2}{5}}(\mathcal{A}_{kn}\partial_\eta u_{g,n}+\mathcal{B}_{kmn}u_{\alpha, m}u_{\chi, n}+J_0\eta \mathcal{J}_k)=0, \nonumber \\
&\lambda_k u_{g,k}+v^{\frac{2}{5}}(\mathcal{A}_{kn}\partial_\eta u_{\chi,n}-\mathcal{B}_{kmn}u_{g,m}u_{\alpha,n})=0.
\ea
Solutions of these EOMs have expansions in terms of small $v$.
From above equations and $\lambda_0=0$, we obtain that the zero mode contributes to $O(1)$, while non-zero modes contribute to $O(v^\frac{2}{5})$. In the very small $v$ limit, the dynamics are described by following sets:
\ba
&\mathcal{A}_{00}\partial_\eta u_{g,0}+\mathcal{B}_{00m}u_{\chi, 0}u_{\alpha, m}+J_0\eta \mathcal{J}_0=0, \nonumber \\
&\mathcal{A}_{00}\partial_\eta u_{\chi,0}-\mathcal{B}_{00m}u_{g,0}u_{\alpha,m}=0,
\ea
where $u_{\alpha, m}$ are obtained from \eqref{EOM68}. Returning to the equation of motion \eqref{EOM11}, the normalizable part of fields should obey following scaling rules:
\ba
w(t,v)=v^{\frac{1}{5}}w(tv^{\frac{2}{5}},1), \quad g(t,v)=v^{\frac{1}{5}}g(t v^{\frac{2}{5}},1),\quad \phi (t,v)=v^{\frac{2}{5}}\phi (tv^{\frac{2}{5}},1).
\ea
This implies that the corresponding VEV has the following Kibble-Zurek scaling:
\ba\label{VEV25}
&\langle O_{w,g}(t,v)\rangle =v^{\frac{1}{5}}F(tv^{\frac{2}{5}}), \\
&\langle \rho (t,v)\rangle =v^{\frac{2}{5}}G( tv^{\frac{2}{5}}).
\ea
These have the same scaling form as obtained in the 4d holographic s-wave superconductor models~\cite{Basu:2012gg,Das:2014lda,Das:2016eao}.
In fact, this is not a coincidence. Recall that both their 4d model and our 2d model have mean field static critical exponents $(\a,\b,\g,\d)=(0,\frac{1}{2},1,3)$. The remaining exponents $\n=\frac{2}{d}$, $\h=2-\frac{d}{2}$ depend on the dimensionality. Furthermore, both models correspond to the dynamical universality model A because the order parameter is non-conserved, thus $z=2-\h=\frac{d}{2}$. It follows that the exponent $\z=\frac{z\n}{\d\b}=\frac{1}{\d\b}$ is independent of the dimensionality!

\section{Summary and Outlook} \label{conclusion}

We have calculated all critical exponents for the (1+1)-d holographic p-wave superconductor. We find the static exponents are of mean field type, and the dynamical exponent corresponds to model A. We have also studied a quench process with a homogeneous source coupled to the order parameter. On general grounds, we are able to express the Kibble-Zurek time scales with the exponent $\z=\frac{z\n}{\d\b}$, which is in fact independent of the dimensionality of the system. We confirm the scaling via holographic analysis of the bulk equation of motion and find the scaling function of the order parameter. The scaling of KZ time and scaling function are formally the same as (3+1)-d s-wave superconductor.

The apparent similarity between s-wave and p-wave models should not be taken too far. The quench we consider is of a very special type, i.e. quench on precisely the critical point by a homogeneous source coupled to the order parameter following an adiabatic time profile. The restriction to the quench can be relaxed in different ways. Firstly, the time profile of the source can be varied. Different protocols of crossing the critical point has been classified in \cite{1202.5277}, which could lead to possible different behavior in the scaling functions. It would be interesting to explore the consequence of different time profile of the source. Secondly, it is of more practical interest to consider an inhomogeneous source, which would lead to defect formation and hydrodynamics. Both are dependent on the dimensionality and symmetry group. Lastly, while the dominance of critical mode in the dynamics is true only when the system is very close to the critical point. Away from the critical point, the dynamics involves both critical mode and hydrodynamic mode. It would be interesting to study the interplay between the two. We leave these for future studies.

\section*{Acknowledgments}

M.F. would like to thank S. R. Das for useful discussions related to this work.
Y.B. is supported by the Fundamental Research Funds for the Central Universities under the grant No. 122050205032 and the Natural Science Foundation of China (NSFC) under the grant No. 11705037. M.F. is supported by the NSFC under the grant No. 11850410431. S.L. is supported by One Thousand Talent Program for Young Scholars and the NSFC under the grant Nos 11675274 and 11735007.

\appendix

\section{An overview of critical exponents} \label{overview_exponents}

In this appendix, we give an overview of critical exponents for self-consistency of this paper. Critical exponents describe the behavior of physical quantities near a continuous (second order) phase transition, such as the liquid-vapour transition on the critical point. It is believed that critical exponents show universal properties of continuous phase transitions. Particularly, they only rely on some of the general features of the physical system, rather than depending on the details of the physical system.

Let us consider a specific continuous phase transition driven by changing the temperature $T$. When $T> T_c$, with $T_c$ the critical temperature where the phase transition occurs, the physical system lives in the highly symmetric phase (or disordered phase). Conversely, when $T<T_c$, the system is in a symmetry-breaking phase (or ordered phase). Around the critical temperature $T_c$, if a physical quantity $\mathcal{A}$ obeys power law behavior,
\begin{align}
\mathcal{A} \propto \epsilon_T^k, \qquad {\rm with} \qquad \epsilon_T= \frac{T-T_c}{T_c},
\end{align}
it then yields a critical exponent $k$. Generally, a continuous phase transition is characterized by an order parameter $\mathcal{O}$, which non-vanishes only when $T<T_c$ if there is no external source $J$. The six static critical exponents $(\alpha,\beta,\gamma,\delta,\nu,\eta)$ are defined as
\begin{align} \label{alpha_gamma_nu}
C_H \propto |\epsilon_T|^{-\alpha}, \quad \chi \propto |\epsilon_T|^{-\gamma},\quad  \xi \propto |\epsilon_T|^{-\nu}, \quad {\rm with} \quad T\neq T_c,
\end{align}
\begin{align} \label{delta_eta}
\mathcal{O}\propto J^{1/\delta},\qquad \langle \mathcal{O}(0) \mathcal{O}(r) \rangle \propto r^{-d+2-\eta},\quad {\rm with} \quad T=T_c,
\end{align}
\begin{align} \label{beta}
\mathcal{O} \propto (-\epsilon_T)^\beta \quad {\rm with} \quad T<T_c \quad {\rm and } \quad J=0,
\end{align}
where $J$ is a possible source for the order parameter $\mathcal{O}$, and $d$ is the spatial dimension. The correlation length $\xi$ is defined as
\begin{align}
\langle \mathcal{O}(0) \mathcal{O}(r) \rangle \propto e^{-r/\xi}, \quad {\rm with} \quad T\neq T_c.
\end{align}
Additionally, the specific heat $C_H$ and the susceptibility $\chi$ are defined as
\begin{align}
C_H \equiv -T \frac{\partial^2\mathcal{F}}{\partial T^2},\qquad \chi \equiv \frac{\partial \mathcal{O}}{\partial J}.
\end{align}
In (\ref{alpha_gamma_nu}), it has been assumed that the critical exponents $(\alpha,\gamma,\nu)$ computed from either high temperature phase ($T>T_c$) or low temperature phase $T<T_c$ are identical. This is indeed true for most cases. For the exponent $\beta$, one has to derive it in the low temperature phase.

From the theory of the renormalization group, the static critical exponents satisfy the following scaling relations:
\begin{align}
&\alpha + 2\beta + \gamma = 2 \qquad ({\rm Rushbrooke}), \label{RUS62} \\
&\gamma = \beta(\delta - 1) \qquad \qquad  ({\rm Widom}); \\
&\gamma = \nu (2 - \eta ) \qquad \qquad  ({\rm Fisher}); \\
&2 - \alpha  = d \nu  \qquad \qquad  ({\rm Josephson}), \label{JOS65}
\end{align}
which imply that there are only two independent exponents among $(\alpha,\beta,\gamma,\delta,\nu,\eta)$. In mean field theory, their values are
\begin{align}\label{ce_mf}
(\alpha,\beta,\gamma,\delta,\nu,\eta)=\left(0, \,  \frac{1}{2} ,\,  1 ,\,  3 ,\, \frac{2}{d} ,\, 2-\frac{2}{d}\right).
\end{align}

In order to further classify the large static universality classes of equivalent models with identical static critical exponents, one needs to introduce dynamical critical exponents \cite{Hohenberg:1977ym}. Of particular interest is the dynamical exponent $z$, which is defined as
\begin{align}
\tau\propto \xi^z,\qquad {\rm as} \quad T\to T_c,
\end{align}
where $\tau$ is the characteristic time of a system, such as relaxation time. The dynamical exponent $z$ is crucial in classifying systems into different dynamical universality classes. If the order parameter does not couple to stress tensor. The system can be classified based on whether the order parameter is conserved (model B) or not (model A). The corresponding dynamical critical exponent is given by:
\begin{equation}
\begin{split}
&{\rm model \quad A}:\qquad z=2-\eta, \\
&{\rm model \quad B}:\qquad z=4-\eta.
\end{split}
\end{equation}

\section{KZ Scaling from the Eddington-Finkelstein coordinates} \label{EF_KZ}

In this appendix, we re-derive the Kibble-Zurek scaling using the ingoing Eddington-Finkelstein (EF) coordinates and demonstrate the same results as \eqref{VEV25}. Moreover, we extend the adiabatic expansion \eqref{adiabatic} to second order.

The metric of the $AdS_3$ black brane is presented in \eqref{btz310} with $\rho=0$.
We set the $AdS$ radius to units where $l=1$. The ingoing EF coordinates $(u,y,z)$ are related to those in \eqref{btz310} by \cite{1109.3909,1308.4061}:
\ba\label{EDD2}
dt=du-\frac{dz}{f(z)},\quad {\rm others~unchanged},
\ea
where $u$ is the time in the ingoing EF coordinates. Note, at the $AdS$ boundary, $u=t$.
In terms of $(u,y,z)$, the line element \eqref{btz310} of the bulk metric becomes
\ba\label{EDD3}
ds^2=2\dfrac{dz du}{z^2}+\dfrac{dy^2}{z^2}-\dfrac{f(z)du^2}{z^2}.
\ea

In the probe limit, we consider the following ansatz in the radial gauge:
\ba\label{BAC4}
A=\dfrac{1}{2}\Big(\phi (u,z)  \sigma^3 du+\vec{w}(u,z) \cdot  \vec{\sigma} dy\Big),\quad A_z^b=0,
\ea
where we switch on two components $w^1\equiv w$ and $w^2 \equiv g$ in $\vec w$.
The field strength with the ansatz \eqref{BAC4} is
\ba
&F_{uy}=\dfrac{\sigma^1}{2}(-\partial_u w+\phi g)+\dfrac{\sigma^2}{2}(\partial_u g+\phi w),\nonumber \\
&F_{yz}=-\Big(\dfrac{\sigma^2}{2}g^{ \prime}+\dfrac{\sigma^1}{2}w^{ \prime}\Big),\quad F_{uz}=\dfrac{\sigma^3}{2}\phi '.
\ea

The non-linear equations of motion in terms of the bulk gauge fields are
\ba\label{EOM16}
&-(z\phi')'-z g^{\prime} w+z g w^{\prime}=0, \nonumber \\
&(zf(z)w^{\prime})' -2 z g^{\prime} \phi- g \left(z \phi^{\prime}+\phi\right)+\pt w+2 z \pt w^{\prime}=0, \nonumber \\
&(zf(z)g^{\prime})' +2 z w^{\prime} \phi+w \left(z \phi^{\prime}+\phi\right)+\pt g+2 z \pt g^{\prime}=0, \nonumber \\
&f(z) (g^{\prime} w-g w^{\prime})+\phi (w^2+g^2) -g \pt w+\pt g w-\pt \phi'=0,
\ea
where the fourth one is the constraint equation. Because EOMs above can be rewritten in terms of the complex field $\Phi\equiv w - ig$,  $w$ and $g$ could be regarded as the real and imaginary parts of $\Phi$, respectively.~\footnote{Using an EF coordinate $Z=\int \frac{dz}{f(z)}$ and rescaling fields like $\Phi=\tilde{\Phi}/\sqrt{z}$, moreover, two EOM can be rewritten like Sturm-Liouville theory. }

Near the $AdS$ boundary, the bulk gauge fields are expanded as
\ba\label{EOM17}
&\phi (u,z) =\phi_1(u)\log z +\phi_0(u)
+ z (g_0(u) w_1(u)-g_1(u) w_0(u))+O(z^2), \nonumber \\
&g(u,z)=g_1(u)\log z +g_0(u) \nonumber \\
&+ z \Big(\log z \Big(-\pt g_1(u)-w_1(u) \phi_0(u)-w_0(u) \phi_1(u)+w_1(u) \phi_1(u)\Big) \nonumber \\
&-\log ^2z w_1(u) \phi_1(u)-w_0(u) \phi_0 (u)+w_0(u) \phi_1(u)-\pt g_0(u)\Big)+O(z^2), \nonumber \\
&w(u,z)= w_1(u)\log z+w_0(u) \nonumber \\
&+ z \Big(\log z \Big(g_1(u) \phi_0 (u)+g_0(u) \phi_1(u)-g_1(u) \phi_1(u)-\pt w_1(u)\Big) \nonumber \\
&+\log ^2z g_1(u) \phi_1 (u)+g_0(u) \phi_0(u)-g_0(u) \phi_1 (u)-\pt w_0(u)\Big)+O(z^2).
\ea
Note the coefficient $\phi_1(u)$ should obey the following constraint:
\ba
\pt \phi _1(u)= g_1(u) w_0(u)-g_0(u) w_1(u).
\ea
%
%
At the $AdS$ boundary, the boundary conditions of the bulk fields are specified by the source terms $\phi_1$, $w_1(u)$ and $g_1(u)$ in \eqref{EOM17}, which will be designed to change in time $u$ slowly.

At the horizon, the bulk gauge fields are required to be regular, which is equivalent to the ingoing wave condition in the Poincare coordinates $(t,y,z)$ of \eqref{btz310}. As a result, near the horizon the gauge fields are expanded as
\ba
&&\phi (u,z)=\phi_{bh,0}(u)+\phi_{bh,1}(u) (1-z)+ \phi_{bh,2}(u) (1-z)^2+O((1-z)^3),\nonumber  \\
&&g (u,z)=g_{bh,0}(u)+g_{bh,1}(u) (1-z)+O((1-z)^2),\nonumber  \\
&&w(u,z)=w_{bh,0}(u)+w_{bh,1}(u) (1-z)+O((1-z)^2),
\ea
where the coefficients satisfy first order (in time $u$) differential equations
\ba\label{DIF19}
&&\phi _{\text{bh},2}(u)= \dfrac{1}{2} \left(g_{\text{bh},1}(u) w_{\text{bh},0}(u)-g_{\text{bh},0}(u) w_{\text{bh},1}(u)+\phi _{\text{bh},1}(u)\right), \nonumber \\
&&\pt \phi _{\text{bh},1}(u)= g_{\text{bh},0}(u)\pt w_{\text{bh},0}(u)-w_{\text{bh},0}(u)\pt g_{\text{bh},0}(u)-g_{\text{bh},0}(u){}^2 \phi _{\text{bh},0}(u)\nonumber \\
&& \qquad \qquad \quad -w_{\text{bh},0}(u){}^2 \phi _{\text{bh},0}(u), \nonumber \\
&&\pt w_{\text{bh},1}(u)= \dfrac{1}{2} \Big(-g_{\text{bh},0}(u) \phi _{\text{bh},0}(u)+2 g_{\text{bh},1}(u) \phi _{\text{bh},0}(u) +g_{\text{bh},0}(u) \phi _{\text{bh},1}(u) \nonumber \\
&& \qquad \qquad \qquad +\pt w_{\text{bh},0}(u)+2 w_{\text{bh},1}(u)\Big), \nonumber \\
&&\pt g_{\text{bh},1}(u)= \dfrac{1}{2} \Big(\pt g_{\text{bh},0}(u)+2 g_{\text{bh},1}(u) +w_{\text{bh},0}(u) \phi _{\text{bh},0}(u)-2 w_{\text{bh},1}(u) \phi _{\text{bh},0}(u)
\nonumber \\
&& \qquad \qquad \qquad -w_{\text{bh},0}(u) \phi _{\text{bh},1}(u)\Big). 
\ea
The time-component of the bulk gauge potential $\phi (u,z)$ usually vanishes at the black hole horizon in a static background, particularly in the Poincare coordinates like \eqref{btz310}. However, it will be nonzero at the horizon in a time-dependent situation, see the QNM analysis or the real-time AdS/CFT~\cite{Policastro:2001yc,Son:2002sd}.


Before solving the bulk equations \eqref{EOM16}, we consider a special case where the time dependence in $\phi,w,g$ is completely turned off. Then, \eqref{EOM16} turns into
\ba \label{no_time_phi_w_g}
&-(z\phi')'-z g^{\prime} w+z g w^{\prime}=0, \nonumber \\
&D_1 w -2z (\phi +\rho \log z)g'-(z\phi'+\phi+\rho+\rho\log z)g=0, \nonumber \\
&D_2 g +2z (\phi +\rho \log z)w'+(z\phi'+\phi+\rho+\rho\log z)w=0, \nonumber \\
&-f(z)(z\phi')'+z\phi (w^2+g^2)=0,
\ea
where
\begin{align} \label{D12}
&D_1 w\equiv (zf(z) w^{\prime})'+2z\rho\log z g^{\prime}+(\rho+\rho\log z)g, \nonumber\\
&D_2 g\equiv (zf(z) g^{\prime})'-2z\rho\log z w^{\prime}-(\rho+\rho\log z)w.
\end{align}
 One can diagonalize the second and third EOM in \eqref{EOM120} by taking a linear combination.
%
Given a time-independent source $J$ when $\rho=\rho_c$, the solutions to \eqref{no_time_phi_w_g} could be written as
\ba\label{FIE5}
w=-J u_{f}^{(0)}(z)+J^{\frac{1}{\delta}}u_{\chi }^{(0)}(z),\quad g=-J u_{f2}^{(0)}(z)+J^{\frac{1}{\delta}}u_{\chi 2}^{(0)}(z), \quad \phi =-\rho \log z+J^v u_{\alpha}^{(0)}(z),
\ea
where we assumed $\delta \ge 1$. If $D_1 u^{(0)}_\chi \neq 0$ and $D_2 u^{(0)}_{\chi 2} \neq 0$, the solution exists when $\delta =1$ and $v=2$.
However, in the presence of a zero mode existing at $\rho =\rho_c$ and satisfying $D_1u_{\chi }^{(0)}=D_2 u_{\chi 2}^{(0)}=0$, the solution satisfies
\ba
v=\dfrac{2}{3},\quad \delta =3.
\ea
In terms of the VEV,
\ba
w_v  \sim J^{\frac{1}{3}},\quad \mu\sim J^{\frac{2}{3}}, \quad
\ea

We turn to solve the bulk equations \eqref{EOM16} given a time-dependent source quench on the critical point $\rho=\rho_c$. Specifically, the time dependent source $J(u)$ changes between two constant values at early and late times as
\ba
J(u)=J_0\tanh (v u), \quad v \ll 1,
\ea
where we have restricted to the slow quench.
The system is adiabatic when $u\to -\infty$, while  the system goes to a critical point for $u\to 0$ and adiabatic approximation breaks down.

As in section \ref{KZ_bulk_eom}, we consider an adiabatic expansion of the bulk fields
\ba\label{ADI6}
&w=w^{(0)}(z,u)+\epsilon u_w^{(1)}(z,u)+\epsilon^2 u_w^{(2)}(z,u)+\dots, \nonumber \\
&g=g^{(0)}(z,u)+\epsilon u_g^{(1)}(z,u)+\epsilon^2 u_g^{(2)}(z,u)+\dots, \nonumber \\
&\phi =\alpha^{(0)}(z,u)+\epsilon u_\alpha^{(1)}(z,u)+\epsilon^2 u_\alpha^{(2)}(z,u)+\dots,
\ea
where $\epsilon$ is an adiabatic parameter, counting the number of time derivative.

At the leading order $O(\epsilon^0)$, there is no time derivative. So, the leading order solutions can obtained by substituting $J \to J(u)$ in \eqref{FIE5}
\ba\label{ANS118}
&&w^{(0)}=-J(u) u_{f}^{(0)}(z)+J^{\frac{1}{\delta}}(u)u_{\chi }^{(0)}(z),\quad g^{(0)}=-J(u) u_{f2}^{(0)}(z)+J^{\frac{1}{\delta}}(u)u_{\chi 2}^{(0)}(z) ,\nonumber \\
&& \alpha^{(0)} =-\rho \log z+J^v(u) u_{\alpha}^{(0)}(z).
\ea

At $O(\epsilon)$, the bulk equations \eqref{EOM16} become
\ba \label{1st_phi_w_g}
&-u_\alpha^{(1)\prime}-z u_\alpha^{(1)\prime\prime}-z g^{(0)\prime} u_w^{(1)}-z u_g^{(1)\prime} w^{(0)}+z u_g^{(1)} w^{(0)\prime}+z g^{(0)} u_w^{(1)\prime}=0, \nonumber \\
&z f'(z) u_w^{(1)\prime}+f(z) u_w^{(1)\prime}+z f(z) u_w^{(1)\prime\prime}-2 z u_\alpha^{ (1)} g^{(0)\prime}-2 z \alpha^{(0)} u_g^{(1)\prime} \nonumber \\
&-\left(\alpha^{(0)}+z \alpha^{(0)\prime}\right) u_g^{(1)}-\left(u_\alpha^{(1)}+z u_\alpha^{(1)\prime}\right) g^{(0)}+\partial_u w^{(0)}+2 z\partial_u  w^{(0)\prime}=0, \nonumber \\
&z f'(z) u_g^{(1)\prime}+f(z) u_g^{(1)\prime}+z f(z) u_g^{(1)\prime\prime}+\partial_u g^{(0)}+2 z \partial_u g^{(0)\prime}+2 zu_\alpha^{(1)} w^{(0)\prime} \nonumber \\
&+2 z \alpha^{(0)} u_w^{(1)\prime}+\left(\alpha^{(0)}+z \alpha^{(0)\prime}\right) u_w^{(1)}+\left(u_\alpha^{(1)}+z u_\alpha^{ (1)\prime}\right) w^{(0)}=0. 
\ea
Recall that the amplitude of the source $J(u)$ becomes small around the critical point $\rho =\rho_c$ with $vt\sim 0$. With the leading order solutions \eqref{ANS118}, the dynamical components in \eqref{1st_phi_w_g} could be expanded in powers of $J(u)$,
\ba\label{EOM120}
&J(u)^{\frac{1}{3}} \Big(z u_\chi^{(0)\prime} u_g^{(1)}- z u_\chi^{(0)} u_g^{(1)\prime}- z u_{\chi 2}^{(0)\prime} u_w^{(1)}+ z u_{\chi 2}^{(0)} u_w^{(1)\prime}\Big)- (z u_{\alpha }^{(1)\prime})'+\dots=0, \nonumber \\
&D_1u_w^{(1)}+J(u)^\frac{1}{3} \Big(-2 z u_{\chi 2}^{(0)\prime} u_{\alpha}^{(1)}-u_{\chi 2}^{(0)} (z u_{\alpha}^{(1)})' \Big) \nonumber \\
&+\dfrac{\pt J(u)}{J(u)^{\frac{2}{3}}}\Big(\dfrac{1}{3} u_{\chi}^{(0)} +\dfrac{2}{3} z  u_\chi^{(0)\prime}\Big)+\dots=0, \nonumber \\
&D_2u_g^{(1)}+J(u)^\frac{1}{3} \Big(2 z u_\chi^{(0)\prime} u_\alpha^{(1)}+u_\chi^{(0)} u_\alpha^{(1)}+z u_\chi^{(0)} u_\alpha^{(1)\prime}\Big) \nonumber \\
&+\dfrac{\pt J(u)}{J(u)^\frac{2}{3}}\Big( \dfrac{1}{3} u_{\chi 2}^{(0)}+\dfrac{2}{3} z u_{\chi 2}^{(0)\prime}\Big)+\dots=0,
\ea
where $D_1,D_2$ are defined in \eqref{D12}. { One can diagonalize the second and third EOM in \eqref{EOM120} by taking a linear combination. }
To solve \eqref{EOM120}, we impose the condition of zero modes $D_1u_w^{(1)}=0$ and $D_2u_g^{(1)}=0$. Recall that $ (z u_{\alpha }^{(1)\prime})'$ does not have a zero mode. Then, from \eqref{EOM120} we conclude the following behaviors for $u_\alpha^{(1)}, u_w^{(1)}, u_g^{(1)}$:
\ba
u_g^{(1)}\sim \dfrac{\dot{J}(u)}{J(u)^{\frac{4}{3}}},\quad u_w^{(1)}\sim \dfrac{\dot{J}(u)}{J(u)^{\frac{4}{3}}},\quad u_\alpha^{(1)}\sim  \dfrac{\dot{J}(u)}{J(u)}.
\ea
The adiabaticity breaks down when $u_w^{(1)},\ u_g^{(1)}\sim J^\frac{1}{3}$ or $u_\alpha^{(1)}\sim J^\frac{2}{3}$. This leads to the Kibble-Zurek time  $t_\KZ\sim v^{-\frac{2}{5}}$. At the Kibble-Zurek time and when adiabaticity breaks down, $u_w^{(1)},u_g^{(1)}\sim v^\frac{1}{5}$ and $u_\alpha^{(1)}\sim v^\frac{2}{5}$.

Likewise, at $O(\epsilon^2)$ the EOMs for $u_\alpha^{(2)}$, $u_w^{(2)}$ and $u_g^{(2)}$ could be expanded in powers of $J(u)$, yielding
\ba\label{EOM122}
&u_{\alpha }^{(2)\prime}\sim -z(u_w^{(1)}u_g^{(1)\prime}-u_w^{(1)\prime}u_g^{(1)}) \sim \dfrac{\dot{J}(u)^2}{J(u)^\frac{8}{3}},
\ea
The equation \eqref{EOM122} demonstrates $u_{\alpha}^{(2)}\sim\dot{J}(u)^2/J(u)^\frac{8}{3}$. At the Kibble-Zurek time and the broken adiabaticity, $u_{\alpha}^{(2)}\sim v^{\frac{2}{5}}$.

\end{document}